\documentstyle[12pt]{article}

\begin{document}

\topmargin 0pt \oddsidemargin 5mm

\setcounter{page}{1} 
\vspace{2cm}

\begin{center}
{\bf Converse coding theorems for quantum source and noisy channel }\\ 
\vspace{5mm} {\large A.E. Allahverdyan, D.B. Saakian }\\ \vspace{5mm} {\em %
Yerevan Physics Institute}\\ {Alikhanian Brothers St.2, Yerevan 375036,
Armenia }\\ saakian @ jerewan1.yerphi.am \\
\end{center}

\vspace{5mm} \centerline{\bf{Abstract}} The weak converse coding theorems
have been proved for the quantum source and channel. The results give the
lower bound for capacity of source and the upper bound for capacity of
channel. The monotonicity of mutual quantum information have also been
proved.

\section{{\bf {Introduction}}}

\vspace{5mm} There is currently much interest in quantum information theory,
the theory of compressing, transmitting and storing of messages, represented
as a some quantum states [1-2,4-12]. The concepts of information can be
properly formulated only in the context of physical theory. Therefore the
physics of information is interestingly from many standpoints, not only
because of its obvious practical and engineering importance. When quantum
effects become important, for example at the level of single electrons and
photons, the existing classical information theory becomes fundamentally
inadequate.\\ In general, quantum information theory contains two distinct
types of problems. The the first type describes transmission of classical
information through a channel (the channel can be noisy or noiseless). In
other words bits encoded as some quantum states and only this states or its
tensor products are transmitted. In the second type of problems an arbitrary
superposition of this states or entanglement states are transmitted. In the
first case the problems can be solved by methods of classical information
theory, at least in principle. But in the second case new physical
representations are needed.\\ Quantum information theory has some
application in the theory of nonideal quantum computers [2]. This
hypothetical computational machines work by laws of quantum mechanics and
can efficiently solve some problems that are believed to be intractable on
any classical computer. However, when we considerate physically realizable
quantum computers, we must consider errors due to the computers-environment
coupling. In other words decoherence and noise penetrate in to the computer.
For this reason at present the formulation of quantum error-correcting codes
is intensively being studied [7-10]. Unlike error-correcting codes the
subject of quantum information theory is not a study of realizable in
practice codes, but investigation of general limits for compression and
transmission rates. It is correct to say that basic research in information
theory is directed towards the understanding of general restrictions of
information compressing and transmitting without taking into account their
practical applications. The general problems of information theory are
coding problems for source (noiseless channel) and noisy channel
[6,7,12,13]. Let us briefly explain what each of them means. A quantum
source is determined by a set of states , each state can be used with
probability $p_i$, $i=1,..,N$. 
\begin{equation}
\label{l1}\rho =\sum_ip_i|\psi _i\rangle \langle \psi _i|,\ \ \sum_ip_i=1.
\end{equation}
We only assume that $\langle \psi _i|\psi _i\rangle =1$ and the states may
be nonorthogonal. There are many quantum sources for fixed $\rho $. We
consider the general quantum evolutions operators $\hat {S_1},\hat {S_2}$
[4,14] as coding and decoding. This operators must be linear, completely
positive and trace-preserving. In this paper we consider only
unit-preserving quantum evolutions operators. 
\begin{equation}
\label{l2}\hat S\rho =\sum_\mu A_\mu ^{\dag }\rho A_\mu ,\ \ \sum_\mu A_\mu
A_\mu ^{\dag }=\hat 1,\ \ \sum_\mu A_\mu ^{\dag }A_\mu =\hat 1,
\end{equation}
where $\{A_\mu \}$ is a some set of operators. In particular $\hat {S_1}$, $%
\hat {S_2}$ may be linear combinations of unitary operations, measurements,
partial traces and others.\\ The problem of coding is generated by a
practical problems, which are connected with compressing of a message or a
data transferring in channel without noise. We define the coding as follows 
\begin{equation}
\label{l3}|\psi _i\rangle \langle \psi _i|\mapsto \hat {S_1}|\psi _i\rangle
\langle \psi _i|=\pi _i,\ \ \rho \mapsto \rho _1=\sum_{i=1}p_i\pi _i.
\end{equation}
And in decoding we have 
\begin{equation}
\label{l4}\pi _i\mapsto \hat {S_2}\pi _i=w_i,\ \ \rho _1\mapsto \rho
_2=\sum_{i=1}p_iw_i.
\end{equation}
Where $\dim \rho =\dim \rho _2\geq \dim \rho _1$. We must get $\hat
{S_1},\hat {S_2}$ with minimal $\dim \rho _1$ for fixed $\rho $ and fidelity 
$F_e$ [3] close to 1 (the degree of closeness is also fixed). 
\begin{equation}
\label{l5}F_e=\sum_{i,j}p_ip_j\langle \psi _i|\hat {S_1}\hat {S_2}(|\psi
_i\rangle \langle \psi _j|)|\psi _j\rangle .
\end{equation}
There are several definitions of fidelity in literature [3,5,12]. This
quantity must characterize the degree of closeness between two density
matrices and equal to 1 if and only if this density matrices are identical.
Author of [4] uses for fidelity another expression without referring to a
concrete representation of $\rho $. 
\begin{equation}
F_e=\langle \psi ^R|\hat {S_1}\hat {S_2}(|\psi ^R\rangle \langle \psi
^R)|\psi ^R\rangle .
\end{equation}
Where $\psi ^R$ is a purification of $\rho $ 
\begin{equation}
|\psi ^R\rangle =\sum_i\sqrt{p_i}|\psi _i\rangle \otimes |\phi _i^R\rangle
,\ \ \langle \phi _j^R|\phi _i^R\rangle =\delta _{ij},
\end{equation}
\begin{equation}
tr_R|\psi ^R\rangle \langle \psi ^R|=\rho ,
\end{equation}
here $\{|\phi _i^R\rangle \}$ is some orthonormal set. The definition is
independent to the concrete choice of this set [1].\\ We pint out that
definition of fidelity in the form [6,13] 
\begin{equation}
\bar F=\sum_ip_i\langle \psi _i|w_i|\psi _i\rangle 
\end{equation}
is nonadequate in general case. As example we consider the fully decoherent
mapping 
\begin{equation}
A_\mu =|\mu \rangle \langle \mu |,\ \ \langle \mu |\acute \mu \rangle
=\delta _{\mu \acute \mu }.
\end{equation}
Now for the density matrix $\sigma $ 
\begin{equation}
\sigma =\sum_\mu c_\mu |\mu \rangle \langle \mu |
\end{equation}
we have $\bar F=1$ but $F_e=\sum_\mu |c_\mu |^2\leq 1$. Indeed in general
case we have [4] 
\begin{equation}
\bar F\geq F_e.
\end{equation}
In other words we may say that $F_e$ taking into account the deformation of
phases for $\psi _i$.\\ We define the rate of coding for source as [3] 
\begin{equation}
R=\log _2\dim \rho _1.
\end{equation}
This quantity characterizes the degree of compressing for $\rho $.\\ Now we
describe the coding problem for a quantum noisy channel, defined by mapping $%
\hat S$. We see that the meaning of a source coding is an extracting of a
redundancy. In noisy channel coding we introduce redundancy, as a result we
have faithful transporting of density matrix $\rho $. Now the $\hat {S_1}$ , 
$\hat {S_2}$ are also coding and decoding procedures. 
\begin{equation}
\rho \mapsto \hat {S_1}\rho \mapsto \hat S\hat {S_1}\rho \mapsto \hat
{S_2}\hat S\hat {S_1}\rho =\rho _2.
\end{equation}
Here $\dim \rho =\dim \rho _2\leq \dim \hat {S_1}\rho $. We must find $\hat
{S_1}$, $\hat {S_2}$ for fixed $\rho $ with fidelity $F_e$ between $\rho $
and $\rho _2$ close to 1.\\ In section 3 we define the problems statements
more precisely and introduce block coding.\\ In this work we get converse
theorems for quantum source and noisy channel. Quantum source were
considered in [6,13]. In [6] the author proved the direct coding theorem for
source. In [13] authors considered a general nonunitary decoding, but used $%
\bar F$ for fidelity. Authors of [6,13] worked with memoryless source. A
quantum channels considered in [6,11].\\ In section 2 the concept of quantum
information is reviewed [5]. Using general properties of relative entropy,
the theorem about monotonicity of the mutual quantum information is proved.
In section 3 our results are discussed.

\section{{\bf {Relative entropy and quantum mutual information}}}

Almost all quantities in quantum information theory are formulated in terms
of von Newmans entropy 
\begin{equation}
S(\rho )=-tr\rho \log _2\rho .
\end{equation}
This quantity characterize the degree of 'unorder' of $\rho $ and invariant
with respect to unitary transformation of $\rho $. The main quantity in
classical information theory is the mutual information between two ensembles
of random variables $X$, $Y$. 
\begin{equation}
I(X,Y)=H(Y)-H(Y/X).
\end{equation}
This is the decrease of entropy of $X$ due to the knowledge of $Y$, and
conversely with interchanging $X$ and $Y$. Here $H(Y)$ and $H(Y/X)$ are
Shannon entropy and conditional entropy [3]. In particular $X$, $Y$ may be
the input and the output of a noisy channel. \\ In 1989 M.Ohya [11] made an
attempt to introduce mutual information in quantum theory. But the mutual
information in quantum theory must take into account a specific character of
quantum information transmission. More reasonable definition of quantum
mutual information was introduced by B.Schumacher and M.P. Nielsen [5] (in
this work authors call this quantity coherent information ). These authors
connected mutual information with the deformation degree of initial density
matrix purification (8,9) 
\begin{equation}
I(\rho ;\hat S)=S(\hat S\rho )-S(\hat 1^R\otimes \hat S(|\psi ^R\rangle
\langle \psi ^R|)),
\end{equation}
\begin{equation}
\hat 1^R\otimes \hat S(|\psi ^R\rangle \langle \psi ^R|))=\sum_{i,j}\sqrt{%
p_ip_j}|\phi _i^R\rangle \langle \phi _j^R|\otimes \hat S(|\psi _i\rangle
\langle \psi _j|).
\end{equation}
Quantum mutual information is the decrease of the entropy after acting of $%
\hat S$ due to the possible distortion of the entanglement state. This value
is not symmetric with respect to interchanging of input and output, and can
be positive, negative or zero (in classical case mutual information is a
nonnegative value ).\\ Quantum relative entropy between two density matrices 
$\rho _1$, $\rho _2$ is defined as follows 
\begin{equation}
\label{a1}S(\rho _1||\rho _2)=tr(\rho _1\log \rho _1-\rho _1\log \rho _2).
\end{equation}
This quantity was introduced by Umegaki [15] and characterizes the degree of
'closeness' of density matrices $\rho _1$, $\rho _2$. The properties of
quantum relative information were reviewed by M.Ohya [11]. Here only one
basic property is mentioned. 
\begin{equation}
\label{a2}S(\rho _1||\rho _2)\geq S(\hat S\rho _1||\hat S\rho _2).
\end{equation}
This inequality was proved by Lindblad [16] and valid not only for
unit-preserving maps, but also in general case. 
As a consequence of Lindblad inequality we can prove the monotonicity of
mutual information [5]. 
\begin{equation}
I(\rho ;\hat {S_1})\geq I(\rho ;\hat {S_2}\hat {S_1}).
\end{equation}
Indeed we have 
\begin{eqnarray}
&   & S(\hat{1}^R\otimes \hat S(|\psi ^R\rangle \langle \psi ^R|)||\hat
{1}^R\otimes \hat S(\rho ^R\otimes \rho  ))\nonumber \\
& =&-S(\hat{1}^R \otimes \hat S(|\psi
^R\rangle \langle \psi ^R|))+S(\rho ^R)+S(\hat S\rho )).
\end{eqnarray}
Here 
\begin{equation}
\rho ^R=\sum_{i,j}\sqrt{p_ip_j}|\phi _i^R\rangle \langle \phi _j^R|\langle
\psi _i|\psi _j\rangle .
\end{equation}
Now from Lindblad inequality we have 
\begin{eqnarray}
&      & S(\hat {1^R}\otimes \hat S(|\psi ^R\rangle \langle \psi ^R|)||\hat
{1^R}\otimes \hat S(\rho ^R\otimes \rho ))\nonumber \\
& \geq & S(\hat {1^R}\otimes \hat
{S_1}\hat {S_2}(|\psi ^R\rangle \langle \psi ^R|)||\hat {1^R}\otimes \hat
{S_1}\hat {S_2}(\rho ^R\otimes \rho )).
\end{eqnarray}
From Lindblad inequality and (2) we have that if $\dim \rho =\dim \hat S\rho 
$ and unit-preserving $\hat S$. 
\begin{equation}
S(\hat S\rho )\geq S(\rho ).
\end{equation}
This inequality is an analog of L.Boltzmans H-theorems.

Further $S(\hat{1^R}\otimes \hat{S} (|\psi ^{R} \rangle \langle \psi ^{R}|))$
will be written as $S(\rho \mapsto \hat{S}\rho)$.\\ In (3) B.Schumacher
proved quantum Fano inequality. 
\begin{equation}
S(\rho \mapsto \hat {S}\rho)\leq (1-F_{e})\log _{2}(d^2-1)+h(F _{e}), 
\end{equation}
\begin{equation}
h(x)=-x\log _{2}x-(1-x)\log _{2}(1-x), 
\end{equation}
where $d=\dim \hat{S} \rho$. In qualitative level the meaning of this
theorem is as follows: If we connect $1- F_{e}$ with probability of error,
then $h(1- F_{e})$ is an information for decision. And if we have error,
then $(1- F_e)\log _{2}(d^2-1)$ is the upper bound of information for
determination of this error.

\section{{\bf {The converse coding theorems}}}

For generality reasons we formulate our theorems in terms of blocks with
length $n$. In other words instead of $\rho $ 
\begin{equation}
\rho ^{(n)}=\rho \otimes ...\otimes \rho 
\end{equation}
is used. The direct theorems of information theory frequently has only
asymptotic character and valid only for $n\gg 1$ [3]. Now all
transformations $\hat {S}_1^{(n)}$, $\hat {S}_2^{(n)}$, $\hat{S}^{(n)}$ acts
on $\rho ^{(n)}$ and if $\hat {S}^{(n)}=\hat{S}\otimes ...\otimes \hat{S} $
then we have memoryless channel (source). We working with any n and with
channel (source) with any memory.

In quantum information theory we have von Newman entropy as an analog of
corresponding quantity $H(X)$, introduced by Shannon [3]. For the source
coding the following theorem is proved: \\ If $R\leq S(\rho )$ then $F_e$
between $\rho ^{(n)} $ and $\rho ^{(n)} _{2}$ is not close to 1 for any n, $%
\hat{S}^{(n)}_1$, $\hat{S}^{(n)}_2$. \\ In other words there exist some $%
\delta $ and $F_e\leq \delta < 1$. In section 4 the problem of existence for
a strong converse theorem in a general case is discussed.\\ With the help of
(15,19) we get 
\begin{equation}
nR \geq S(\rho _{1}^{(n)})\geq I(\rho ^{(n)}; \hat {S}^{(n)}_{1}). 
\end{equation}
From the (23) we obtain the following formula 
\begin{equation}
nR\geq I(\rho ^{(n)}; \hat {S}^{(n)}_{2}\hat {S}^{(n)}_{1})= S(\rho
_{2}^{(n)})- S(\rho ^{(n)}\mapsto \rho _{2}^{(n)}). 
\end{equation}
With the help of H-theorem we come to 
\begin{equation}
nR\geq nS(\rho )-S(\rho ^{(n)}\mapsto \rho ^{(n)}_{2}). 
\end{equation}
Now if $R=S(\rho )-\delta$, with $\delta \geq 0$ from quantum Fano
inequality we have 
\begin{equation}
\delta \leq ((1-F_{e})\log _{2}(d^{2n}-1)+h(F _{e}))/n. 
\end{equation}
We see that for any n, $F_e$ is not close to 1. If $n \rightarrow \infty $
then 
\begin{equation}
\delta \leq 2(1-F_{e})\log _{2}d. 
\end{equation}
Capacity $C_s$ of a quantum source is defined as follows.\\ If $R\geq C_s$
then there exist some $\hat {S}^{(n)}_{2}$, $\hat {S}^{(n)}_{1}$ with $F_e$
is close to 1. If $R\leq C_s$ then for any $\hat {S}^{(n)}_{2}$, $\hat{S}%
^{(n)}_{1}$ $F_e$ is not close to 1.\\ In [6] B.Schumacher proved existence
of $\hat{S}^{(n)}_{2}$, $\hat{S}^{(n)}_{1}$ with $\bar{F}$ close to 1 and $%
\log _{2}\dim \rho _{1}=nS(\rho )$ for large n. Our inequality show, that
for sufficiently general coding and decoding schemes and fidelity $F_{e}$ we
have $C_{s}\geq S(\rho )$.\\ Now we pass to the converse theorem for noisy
quantum channel. \\ In a channel coding we introduce redundancy in the step $%
\rho \mapsto \rho ^{(n)}$ (in source coding problem we directly work with $%
\rho ^{(n)}$). After this $\hat{S}^{(n)}_{1}$ is a unitary transformation.
The concrete form of this transformation depend from concrete coding schemes
[8-10]. Decoding transformation $\hat{S}^{(n)}_{1}$ may be nonunitary,
because it contain a determination of unknown quantum state. In other words
in channel coding $\hat{S}^{(n)}_{1}$ is a unitary transformation, $\hat{S}%
^{(n)}_{2}$ is a general but unit-preserving. We introduce the quantity $%
\tilde{C}$ as 
\begin{equation}
\tilde{C}=\max_{\{ p_i\} } I(\rho ^{(n)} ;\hat {S}^{(n)})/n. 
\end{equation}
Here $\hat{S}^{(n)}$ is a quantum channel. We define the rate of noisy
channel coding as 
\begin{equation}
R_{c}=\max_{\{ p_i\} }. S(\rho ) 
\end{equation}
If $\langle \psi _{i}|\psi _{j}\rangle =\delta _{ij}$ we have a
classical-like input ensemble of random variables, and $S(\rho ^{n})$ is
maximized by the uniform distribution of $p_i$ and the famous classical
expression for $R_{c}$ [3] is obtained 
\begin{equation}
R_{c}=\log _{2}d. 
\end{equation}
For channel coding we will prove the following theorem. \\ If $R_{c}\geq 
\tilde{C}$, then $F_e$ (between $\rho ^{(n)} $ and $\rho ^{(n)} _{2}$) is
not close to 1 for any n, $\hat{S}^{(n)}_1$, $\hat{S}^{(n)}_2$. \\ In other,
words exist some $\delta $, that $F_e\leq \delta < 1$. \\ Derivation of this
theorem is similar to source coding theorem. We have from (23) 
\begin{equation}
I(\rho ^{(n)} ;\hat{S}^{(n)})\geq I(\rho ^{(n)} ;\hat{S}^{(n)}_{2} \hat{S}%
^{(n)} \hat{S}^{(n)}_{1}). 
\end{equation}
Now with $\tilde{C}$ we get to the following chain of inequalities for any
set $\{ p_{i}\} $ 
\begin{eqnarray}
n\tilde{C} & \geq & I(\rho ^{(n)} ;\hat{S}^{(n)}) \nonumber \\
 & \geq & I(\rho ^{(n)} ;\hat{S}^{(n)}_{2} \hat{S}^{(n)} 
\hat{S}^{(n)}_{1}) \nonumber \\
& \geq &  S(\rho ^{(n)}_{2})
-  S(\rho ^{(n)} \mapsto 
\rho ^{(n)}_{2}), 
\end{eqnarray}
\begin{eqnarray}
n\tilde{C} & \geq & S(\rho ^{(n)})-  S(\rho ^{(n)} 
\mapsto \rho ^{(n)}_{2})\nonumber \\
& = & nS(\rho)-  S(\rho ^{(n)}
\mapsto \rho ^{(n)}_{2}). 
\end{eqnarray}
Now we may work with the set $\{{p_i} \} $, which maximized input entropy
(36) 
\begin{eqnarray}
n\tilde{C} & \geq & nR_{c}-S(...\mapsto ...) \nonumber \\
S(...\mapsto ...) & \geq & (R_c-\tilde{C})n \nonumber \\
(R_c-\tilde{C})n & \leq & 
((1-F_{e})\log _{2}(d^{2n}-1)+h(F _{e}))/n.
\end{eqnarray}
Capacity $C_c$ of a quantum channel is defined as following.\\ If $R\leq C_c$
then for channel $\hat{S}^{(n)}$ exists some coding and decoding procedures $%
\hat {S}^{(n)}_{2}$, $\hat {S}^{(n)}_{1}$ with $F_e$ between input and
output is close to 1.\\ If $R\geq C_s$ then for any $\hat {S}^{(n)}_{2}$, $%
\hat {S}^{(n)}_{1}$ $F_e$ is not close to 1.\\ We see, that indeed $\tilde{C}
\geq C_{c}$. The question about direct coding theorem for some restricted
class of channels in quantum information theory is still open.

\section{{\bf {Conclusion}}}

Recently in literature the question about capacity of a quantum channel was
discussed. Quantum Fano theorem was proved in [3]. But Fano theorem is not a
general converse theorem. Fano-like theorem may be written for several
independent quantities [11]. As it has been mentioned, the authors of [12]
considered the capacity of quantum source for general decoding mapping, but
with nonadequate fidelity.\\ Our results suggest that in sufficiently
general case the capacity of a quantum source indeed equal $S(\rho )$. 
The upper bound for the capacity of quantum noisy channel has been found.
This quantity is less than other candidates for a quantum channel capacities
role [12]. \\ In general case the proposition of the weak converse theorem
cannot be strengthened [3]. Strong converse theorem has the following
structure.\\ If $R\geq \tilde C$ is true for quantum channel ($R\leq S(\rho )
$ for source) then for $n\gg 1$ and any $\hat {S_1}$, $\hat {S_2}$ $F_e$ is
close to 0. \\ In classical theory [3] strong converse theorems can be
proved only for simple memoryless (may be in some effective sense, see [3])
channels or sources. A proof of such theorems in quantum theory for general
coding and decoding schemes is still an open problem. In the [7] author in
qualitative manner considered the direct coding problem for memoryless
channel. He gave some arguments that $\tilde C\leq C_c$, but did not
consider the entanglement fidelity.

\end{document}